\documentclass[twocolumn,showpacs,preprintnumbers,amsmath,amssymb]{revtex4}
\usepackage{graphicx}
\usepackage{dcolumn}
\usepackage{bm}
\usepackage[tight]{subfigure}

\newcommand{\half}{\textstyle \frac{1}{2}}
\newcommand{\mvec}[1]{\bm{#1}}

\begin{document}
\title{
  Vibrational detection and control of spin
  in mixed-valence molecular transistors
}
\author{F. Reckermann$^{(1,2)}$}
\author{M. Leijnse$^{(1)}$}
\author{M. R. Wegewijs$^{(1,2)}$}
\affiliation{
  (1) Institut f\"ur Theoretische Physik A, RWTH Aachen, 52056 Aachen,  Germany \\
  (2) Institut f\"ur Festk{\"o}rper-Forschung, Forschungszentrum J{\"u}lich, 52425 J{\"u}lich,  Germany
}
\begin{abstract}
  We investigate electron transport through a mixed-valence molecular complex in which an excess electron can tunnel between hetero-valent transition-metal ions, each having a fixed localized spin. We show that in this class of molecules the interplay of the spins and the vibrational breathing modes of the ionic ligand-shells allows the total molecular spin to be detected as well as controlled by non-equilibrium transport. Due to a \textit{spin-dependent pseudo Jahn-Teller effect} electronic transitions with different spin values can be distinguished by their vibronic conductance side-peaks, without using an external magnetic field. Conversely, we show that the spin state of the entire molecule can also be controlled via the non-equilibrium quantized molecular vibrations due to a novel \textit{vibration-induced spin-blockade}.
\end{abstract}

\pacs{
  73.63.-b,
  85.65.+h,  
  71.70.Ej,    
  85.75.-d    
}

\maketitle
\section{Introduction.}
In recent years,
major experimental advances have been made in contacting and measuring single 
molecules in three-terminal transport junctions~\cite{Park99elmig},
detecting the
vibration~\cite{Park00,Pasupathy04,Osorio07b},
spin~\cite{Osorio07b}
and magnetic properties~\cite{Heersche06,Jo06}.
Using the electrical gate to control the transport
one can perform a substantial analysis of the transport processes,
even when detailed microscopic information about the junction is lacking.
Single-electron and coherent multi-electron tunneling processes
(cotunneling, Kondo effect) governed by strong Coulomb
and electron-vibration interactions
allow detailed information to be extracted,
as demonstrated recently for a single oligo-phenelyne molecule~\cite{Osorio07a,Osorio07b}.
More detailed information can be accessed
when mechanical control of a gateable junction~\cite{Champagne05}
is possible.
In the light of this progress of experimental possibilities
the interesting question arises
how quantum mechanical states
involving both electronic and mechanical degrees of freedom may be detected and, 
perhaps, controlled in transport measurements.
A particularly interesting aspect of single-molecule devices
is the strong coupling of the electron current to the mechanical motion
and the fully quantum mechanical character of this motion,
resulting in Franck-Condon (FC) resonances~\cite{Flensberg03,Mitra04b}.
In addition, in molecular systems with magnetic ions,
the spin degree of freedom becomes important~\cite{Heersche06,Jo06,Romeike06a,Romeike06b,Romeike06c,Grose08}.
Thus interesting magnetic electro-mechanical effects are to be expected~\cite{Fedorets05,Cornaglia05b}.
\par
\begin{figure}
  \includegraphics[width=7cm]{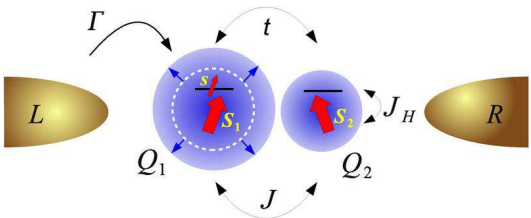}
  \caption{
    Sketch: Mixed-valence dimer trapped between two electrodes, $L,R$.
    By applying a bias voltage, $V_b = \mu_L-\mu_R$,
    electrons can tunnel from one electrode to the other via the molecule.
    A third gate terminal (not shown), coupled capacitively to the molecule,
    effectively shifts the molecular energy levels.
  }
  \label{fig:setup}
\end{figure}

Mixed-valence molecules exhibit this interplay of quantum
nanomechanics and spin-tronics. They are crucial also as a building
block for supra-molecular devices and serve as benchmark for such systems.
In a mixed-valence dimer, sketched in Fig.~\ref{fig:setup},
an excess electron can be localized on either of two equivalent metal ions with a local spin,
thereby locally distorting the positions of the ligand atoms coordinating the ion.
As the electron becomes delocalized over the molecule
the distortion is ``dragged'' along coherently.
As discussed in detail by Bersuker and Borshch this results in vibronic mixing i.e.
the molecular eigenstates do not allow an adiabatic Born-Oppenheimer (BO) separation of the nuclear and electronic motions~\cite{Bersuker92}.
The electronic and vibrational degrees of freedom become entangled into \emph{vibronic states}
due to a \emph{spin-dependent} pseudo Jahn-Teller (pJT) effect.
In general, pJT mixing of electronic and vibrational degrees of freedom
cannot be neglected whenever electronic energy surfaces come close in energy
(not necessarily degenerate as in the Jahn-Teller effect).
The pronounced dependence of the delocalization, and hence of the pJT mixing,
on the total spin of the molecule
arises due to local direct exchange interaction on the ions (Hund's rule),
which favors the excess electron spin to be aligned with the ionic spin.
Simultaneously, in such molecules this spin-dependent kinetic energy gain
is responsible for the ferromagnetic double-exchange interaction~\cite{Anderson55}
which competes with other types of exchange interaction.
\par
In this paper, we present transport calculations for a
model representative of a class of mixed-valence molecules.
We demonstrate that
a single-electron transport current can both detect and control the
molecular spin due to the non-equilibrium nature of the vibrational motion.
As indicated above, this does not rely on weak spin-orbit effects,
but rather on strong direct, kinetic and double-exchange mechanisms.
The pseudo Jahn-Teller dynamics shows up in pronounced sets of vibronic conductance peaks
which depend on the spin-values of the molecular excitations.
This provides a way to detect the spin without a magnetic field
and probe the \emph{in-situ} properties of a mixed-valence molecular transistor.
Conversely, we show that the electronic transport current
induces non-equilibrium quantized molecular vibrations
which drive a pronounced population inversion among the spin-states.
Such a molecule can thereby be switched to a state
with a well-defined charge and spin
by adjusting the applied voltages.
This vibration-induced spin control arises from the interplay of spin
and vibrations intrinsic to mixed-valence molecules,
which may open up new possibilities for detection of mechanical motion
and dissipation, magnetic switching and bistability.

\section{Model.}
We study the non-equilibrium transport properties of a
well-established model for mixed-valence dimers from chemical physics,
see~\cite{Bersuker96,Bersuker92} and the references therein for detailed
discussions.
The model describes the simplest type of molecule exhibiting the spin-vibration interplay,
consisting of two identical transition-metal ions, labeled by
$i=1,2$.
We account for orbital-degenerate electronic states, $\vert i \sigma\rangle$,
where an excess electron with spin projection $\sigma$ is localized on ion $i$.
Within the molecule, the electron can tunnel with amplitude $t$
between the ions via a mechanically stiff bridging ligand.
In addition,
the atomic groups that form a ligand-shell around each ion can vibrate along
the local totally symmetric nuclear coordinate $Q_i$ (breathing mode).
In the absence of an excess electron, this vibration around $Q_i=0$
is assumed to be harmonic with frequency $\omega$.
The excess electron, however, distorts the ligand-shell of the ion on which it resides,
leading to a significant shift of the potential minimum of this mode
(change in bond-lengths).
In units of the zero-point motion energy of the vibration,
this shift is equal to $\sqrt{2}\lambda$, where
$\lambda$ denotes the dimensionless electron-vibration coupling.
In line with many experimental findings,
we assume charging effects (Coulomb blockade) to be strong enough
that only two molecular 
charge states need to be accounted for.
The vibrational parts of the Hamiltonians $H_{\text{vib}}^N$
for $N=0,1$ excess electrons on the molecule,
written in the molecular vibrational coordinates $Q_\pm=(Q_2 \pm Q_1)/\sqrt{2}$, read
\begin{eqnarray}
  H_\text{vib}^0&=&\sum_{i=\pm} \half \omega \left( P_i^2+Q_i^2\right),
		 \label{eq:H_vib} \\
  H_\text{vib}^1&=& H_\text{vib}^0
                   - \lambda\omega Q_+
                   + \lambda\omega Q_- (\hat{n}_{1}-\hat{n}_{2}) \nonumber \\
	        & &+ t \sum_{\sigma} ( d_{1\sigma}^\dagger d_{2\sigma} +h.c.),
		 \label{eq:H_vib_pjt}
\end{eqnarray}
where $\hat{n}_i=\sum_{\sigma}d_{i\sigma}^\dagger d_{i\sigma}$
is the occupation operator of ion $i$ and
$d_{i\sigma}^\dagger$ creates an electron in state $\vert i \sigma \rangle$.
The symmetric coordinate, $Q_{+}$, is the \emph{molecular} breathing mode,
which couples to the total excess charge, $N$, of the molecule,
resulting in a shift by $\lambda$ of its potential surface along $Q_{+}$.
The resulting FC transport effects have been investigated theoretically~\cite{Braig03a,Mitra04b,Koch04b,Wegewijs05}
and found experimentally~\cite{Park00,Pasupathy04,Osorio07a}.
In contrast, the anti-symmetric mode, $Q_{-}$,
couples to the internal charge imbalance $\hat{n}_{1}-\hat{n}_{2}$.
Together with the intra-molecular tunneling, $t$,
this results in the pseudo Jahn-Teller (pJT) effect~\cite{Bersuker92},
i.e. the Hamiltonian~(\ref{eq:H_vib_pjt}) mixes electronic
and vibrational states of the mode $Q_{-}$.
\par
One of the crucial aspects in a mixed-valence molecular transistor
is that the non-trivial dynamics of the $Q_{-}$ mode
depends on the relative orientation of the local transition-metal ion spins.
The molecular Hamiltonian for $N=0,1$ excess electrons reads
\begin{eqnarray}
  H^{N}&=&H_\text{vib}^{N}  - J \mvec{S}_1\cdot\mvec{S}_2\; \nonumber \\
       & &  - J ( \mvec{s}_1 \cdot\mvec{S}_2+\mvec{S}_1\cdot \mvec{s}_2 )
             -\sum_{i=1,2} J_H \mvec{S}_i\cdot\mvec{s}_i.
  \label{eq:H_MV}
\end{eqnarray}
The intra-ionic Hund interaction, $J_H$, couples
the spin of the excess electron,
$\mvec{s}_i=\half\sum_{\sigma,\sigma'}d_{i\sigma}^\dagger \bm{\sigma}_{\sigma\sigma'} d_{i\sigma'}$,
to the spin of the transition-metal ions, $\mvec{S}_i$,
where $\mvec{\sigma}$ denotes the vector of Pauli-matrices
 and $\mvec{s}_i=0$ if no electron is present on ion $i$.
Together with the intra-molecular tunneling, $t$, incorporated in Eq.~(\ref{eq:H_vib_pjt}),
the intra-ionic Hund interaction results in a double-exchange splitting
$2t_S$ of eigenstates with total molecular spin $S$~\cite{Anderson55}:
\begin{eqnarray}
  \frac{t_S}{t}&=&\frac{S+\half}{2S_1+1}\leq 1
  , \label{eq:tS}
\end{eqnarray}
where $S_1=S_2$ denotes the spin-length of the equivalent ions ($\mvec{S}_1,\mvec{S}_2$).
Result~(\ref{eq:tS}) for the effective  tunneling strength $t_S$ is obtained
by expressing $H^1$ in the electronic basis of total spin eigenstates using vector coupling coefficients.
In the semi-classical limit of large ionic spins~\cite{Anderson55}, $S_1 \gg 1$ (not considered further below),
this reduces to ${t_S}/{t}={S}/{2S_1}=\text{cos}(\theta/2)$,
where $\theta$ is the angle between the two classical ionic spins.
This makes clear that the kinetic energy which can be gained by the excess electron
is maximal for parallel ionic spins due to the strong intra-ionic coupling
and is suppressed by $\text{cos}(\theta/2)$, i.e. by the
electron spin-eigenfunction component
quantized in the direction of  the ionic spin.
Importantly, in mixed-valence molecules $J_H$ is much larger than the other energy scales,
$J_H \gg |J|,\omega, t$
(typical values: $J_H \sim$~1~eV, $J,t \sim$~1-100 meV, $\omega \sim $~tens of meV)~\cite{Bersuker92,Piepho78,Borshch91}.
This scale separation derives from the intra-\textit{ionic} origin of
$J_H$ (direct Hund exchange), in contrast to the intra-\textit{molecular}
processes involved in the exchange $J$ and hopping $t$.
We assume a ferromagnetic coupling, $J_H>0$, i.e. a less than half filled ionic shell.
The excitations where $\mvec{S}_i$ and $\mvec{s}_i$ are aligned anti-parallel can therefore be neglected.
Eq.~(\ref{eq:H_MV}) also incorporates the intra-molecular coupling $J$ of the spins of different ions.
From hereon we take the length of the ionic spins to be $S_1=S_2=1/2$.
The Hamiltonian for the charged molecule consists of an $S=3/2$ and an $S=1/2$ diagonal block,
with $2S+1$ sub-blocks on the diagonal, given by
\begin{eqnarray}
  H^1_S&=& H_{\text{vib}}^1|_{t=t_S} - \half J S(S+1) + \text{const.}
  \label{eq:H_MV_blocks}
\end{eqnarray}
This makes explicit the interesting property of mixed-valence molecules,
that the strength of the pJT effect
depends on the total molecular spin
$S$~\cite{Bersuker92}
due to the competition between the local distortion (coupling $\lambda\omega$)
and the effective delocalization of the electron (energy $t_S$).
For $N=1$ we numerically diagonalize the blocks separately for each spin state $S$ to obtain the molecular eigenstates,
which are not of the adiabatic BO form.
For $N=0$ the states trivially factorize in BO form.
\par
\section{Transport.}
The interplay of vibrational and spin degrees of freedom can be demonstrated for the basic transistor setup,
sketched in Fig.~\ref{fig:setup}, described by the transport Hamiltonian
$H = H_\text{MV} + H_\text{T}+H_\text{res}$.
Here $H_\text{MV}=\sum_{N=0,1} X^N H^N X^N - \alpha V_g N$, where $X^N$ projects onto
states with charge $N$ and  $\alpha$ is the gate coupling.
The electrodes $r=L,R$ at electrochemical potential $\mu_r=\mu\pm V_b/2$ and temperature $T$
are described by
\begin{eqnarray}
  \label{eq:ham_res}
  H_\text{res} &=& \sum_{r=L,R}\sum_{k,\sigma} (\epsilon_{k}-\mu_r)c_{rk\sigma}^\dagger c_{rk\sigma},
\end{eqnarray}
i.e. $V_b$ is the bias voltage, and couple to the molecule by
\begin{eqnarray}
  \label{eq:ham_tun}
  H_\text{T} &=&  \sum_{r=L,R}\sum_{i=1,2}\sum_{k,\sigma} T_{r}^i d_{i\sigma}^\dagger c_{rk\sigma} + h.c..
 \end{eqnarray}
Here $c_{rk\sigma}^\dagger$ creates an electron with spin $\sigma$ in state $k$
of electrode $r$ and
$T_r^i$ is the amplitude for tunneling to ion $i$ of the dimer.
We assume a sequential  arrangement  i.e.
$T_L^1=T_R^2= \sqrt{\Gamma / (2\pi \rho)}$ and $T_L^2=T_R^1=0$,
which may experimentally be favored by appropriate design of
``clipping'' ligands.
Here $\Gamma$ denotes the tunneling rate.
\par
In three-terminal molecular junctions, the tunneling rate can be as
small as $\Gamma \approx 0.1$~meV and is often the smallest energy
scale in the problem,
enabling accurate spectroscopy of e.g. molecular vibrations~\cite{Osorio07a}.
We focus on the regime of voltages and temperatures ($T \approx 10\Gamma$)
where single-electron tunneling dominates the transport
as is the case in many experiments~\cite{Park00,Pasupathy04,Heersche06,Jo06,Osorio07a,Osorio07b}.
Using a kinetic (master) equation we calculate the non-equilibrium stationary state occupations of the molecule
for each charge ($N$) and spin multiplet ($S$),
keeping track of the symmetric vibrational ($Q_{+}$) quantum number as well as
the quantum number for the entangled state of the electrons and the anti-symmetric vibrations ($Q_{-}$).
For  weak tunnel coupling the transport rates can be evaluated in lowest non-vanishing order in $H_T$ (i.e. Fermi's Golden Rule).
Additionally, we account phenomenologically for relaxation due to
coupling to a dissipative environment (e.g. substrate phonons).
We assume an energy-dependent density of states
(rate $\gamma^{\text{s(v)}}(E)=\gamma^{\text{s(v)}}_0\cdot(E^2/\omega^2)$)
for transitions between states with equal spin ($\gamma^{\text{v}}_0$)
and different spin ($\gamma^{\text{s}}_0$).
The latter relate to spin-orbit coupling effects and are therefore
assumed to be much smaller than the former, $\gamma^{\text{s}}_0 \ll \gamma^{\text{v}}_0$.
The strength of the spin-allowed relaxation of course depends on the type of the molecular vibration mode
and the junction substrate. For examples and discussion of very slow relaxation in the context of photon-tunneling
in single-molecule junctions, see~\cite{Volokitin07} and the references therein.
\begin{figure}[htbp]
  \centering
  \begin{tabular}{c}
    \begin{tabular}{c}
      \includegraphics[width=8.5cm]{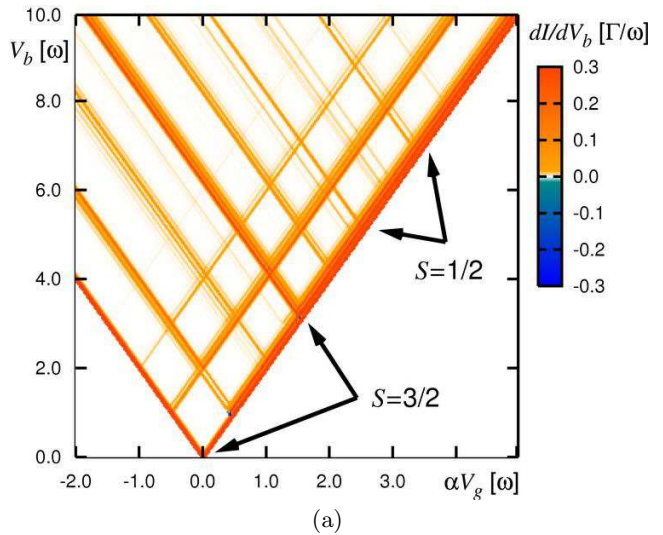}\\
      (a)
    \end{tabular}\\
    \vspace{0.5cm}\\
    \begin{tabular}{c}
      \includegraphics[width=5.5cm]{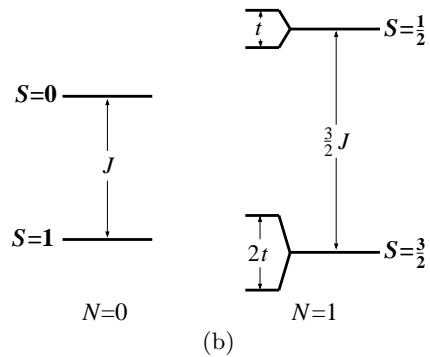}\\
      (b)
    \end{tabular}
  \end{tabular}
  \caption{
    (color online)
    (a) Differential conductance ($dI/dV_b$) for $V_g$ vs. $V_b$ ($J=2.9\omega$, $t=1.5\omega$, $\lambda=0.5$, $\Gamma=9\cdot10^{-4}\omega$, $\gamma^{\text{v}}_0=3.2\cdot10^{-3}\Gamma$, $\gamma^{\text{s}}_0=10^{-4}\Gamma$,  $T=10^{-2}\omega$)
    (red: $dI/dV_b>0$, blue: $dI/dV_b<0$).
    The double-exchange coupling leads to a spin-dependent gap size
    of the vibronic spectrum (see arrows).
    The spectrum of $S=3/2$ is harmonic (signalled by equidistant
    resonance lines of small energy separation),
    while the one of $S=1/2$ is anharmonic (non-equidistant lines).
    (b) Sketch of energy spectrum:
    The spin-multiplets are split due to the intra-molecular coupling $J$.
    For $N=1$, each spin-multiplet is split again by approximately
    twice the spin-dependent intra-molecular tunneling, $t_S$,
    due to the pJT effect.
    Vibrational / vibronic excitations are omitted for clarity.
  }
  \label{fig:result_pJT}
\end{figure}\\
\section{Results.}
\subsection{Spin-dependent pseudo Jahn-Teller effect: Identifying spin values.}
The differential conductance as function of the applied voltages is shown in Fig.~\ref{fig:result_pJT}(a),
using a set of parameters representative for mixed-valence dimers with
weak electron-vibration coupling and ferromagnetic intra-molecular coupling:
$J=2.9\omega$, $t=1.5\omega$, $\lambda=0.5$.
The transport spectrum displays a number of sharp, well-separated resonance lines, which are ``dressed''
by many more lines with small separations.
Two pronounced pairs of excitations appear, corresponding to the $S=1/2$ and $S=3/2$ spin-multiplets
which are split due to the intra-molecular coupling $J$ (c.f. Eq.~(\ref{eq:H_MV_blocks})).
Each multiplet is split approximately by twice the spin-dependent intra-molecular tunneling $t_S$,
leading to a double-exchange gap which is reduced by a factor 2 for the $S=1/2$ state ($t_{1/2}=t_{3/2}/2$, see also Fig.~\ref{fig:result_pJT}(b)).
More generally, the spin parameters
can be determined from the ratios of the
intra-molecular splittings giving $t_{S}/t_{S-1}=(S+1/2)/(S-1/2)$.
A central result of this work is that  an independent check of this assignment of the spin is provided by the vibronic lines 
``dressing'' the above excitations.
This type of excitations arises when the \emph{shape} of the vibrational potentials has a significant charge dependence~\cite{Wegewijs05}.
Here their occurrence indicates a significant pJT mixing in the $N=1$ charge state
which changes the frequency and additionally induces anharmonicity in the effective adiabatic potentials.
If the pJT effect is weak (as for $S=3/2$),
these potentials are approximately harmonic in both charge states, 
and the spacing between these lines (that are due to transitions between excited vibrational and vibronic states)
is even and equals the small frequency \emph{difference}.
For a stronger pJT effect ($S=1/2$)
the potential in the $N=1$ charge state becomes anharmonic and the lines are unevenly spaced.
Clearly, in Fig.~\ref{fig:result_pJT}(a) the ''dressing'' of the lower pair of lines is more evenly spaced
than the upper set of lines, confirming the assignment of high spin state at low energy.
Here we merely note that
a detailed analysis of the pJT transport resonances allows the electro-mechanical parameters $t_{S}, \lambda, \omega$ to be determined quantitatively,
by reading off the voltage positions of the resonances.
Also, we have invoked an adiabatic picture for the interpretation, which has only an approximate validity,
and some qualitative transport effects violate it.
This is, however, not essential here, see~\cite{Reckermann08b} for details.
The spin identification works very well for $t\gtrsim \lambda^2\omega$,
i.e. when the pJT effect is weak to moderate
(for $|J| \gtrsim t$ the resonances of the two spin-multiplets can
be considered as separate).
Thus the intra-molecular ferromagnetic coupling is revealed by the transport spectrum at zero magnetic field,
by double-exchange and vibronic effects.
Finally, we note that the energy average of the total $S$ multiplets split by double-exchange follows $J\,S$,
providing a third, independent check of the spin value assignment.
\subsection{Current blockade and spin switching.}
\begin{figure}[htbp]
  \centering
  \begin{tabular}{c}
    \begin{tabular}{c}
      \includegraphics[width=8.5cm]{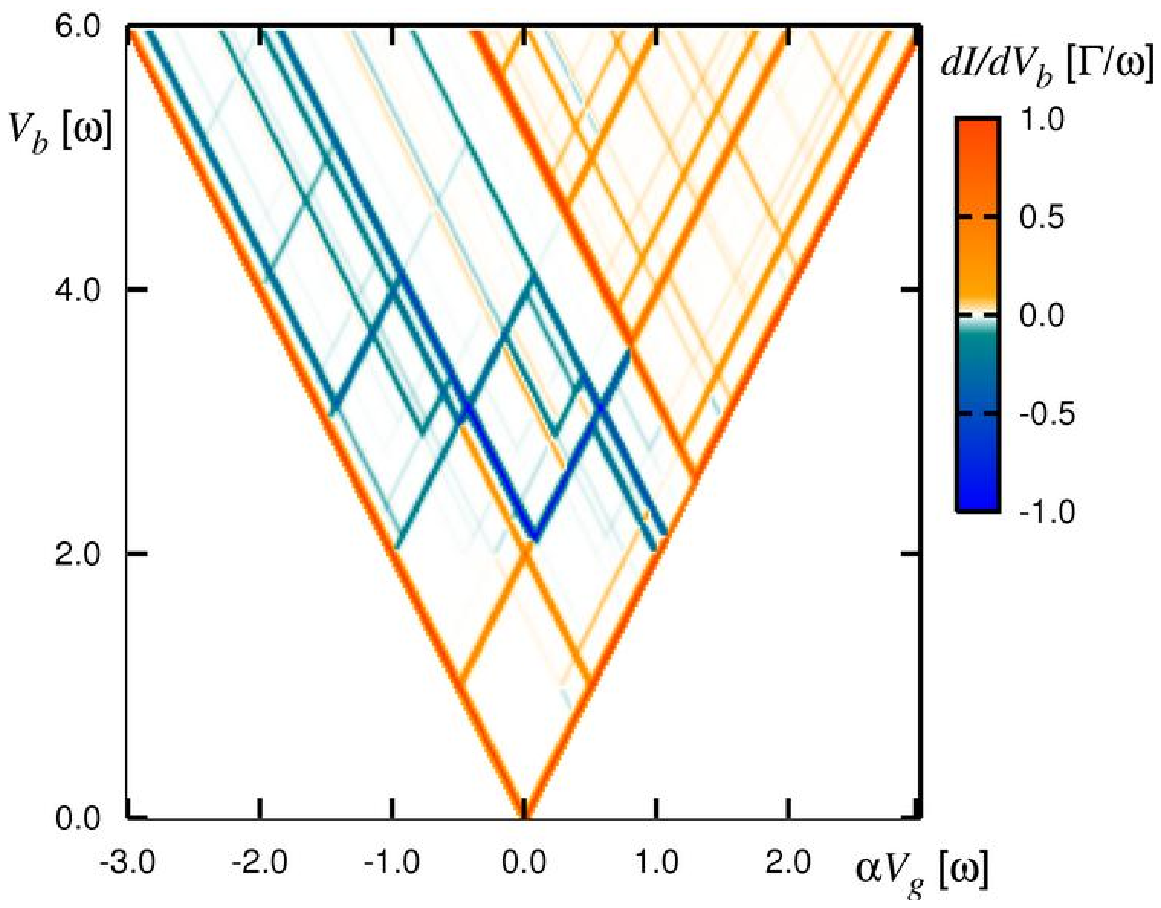}\\
      (a)
    \end{tabular}
    \\
    \begin{tabular}{c}
      \includegraphics[width=6.0cm]{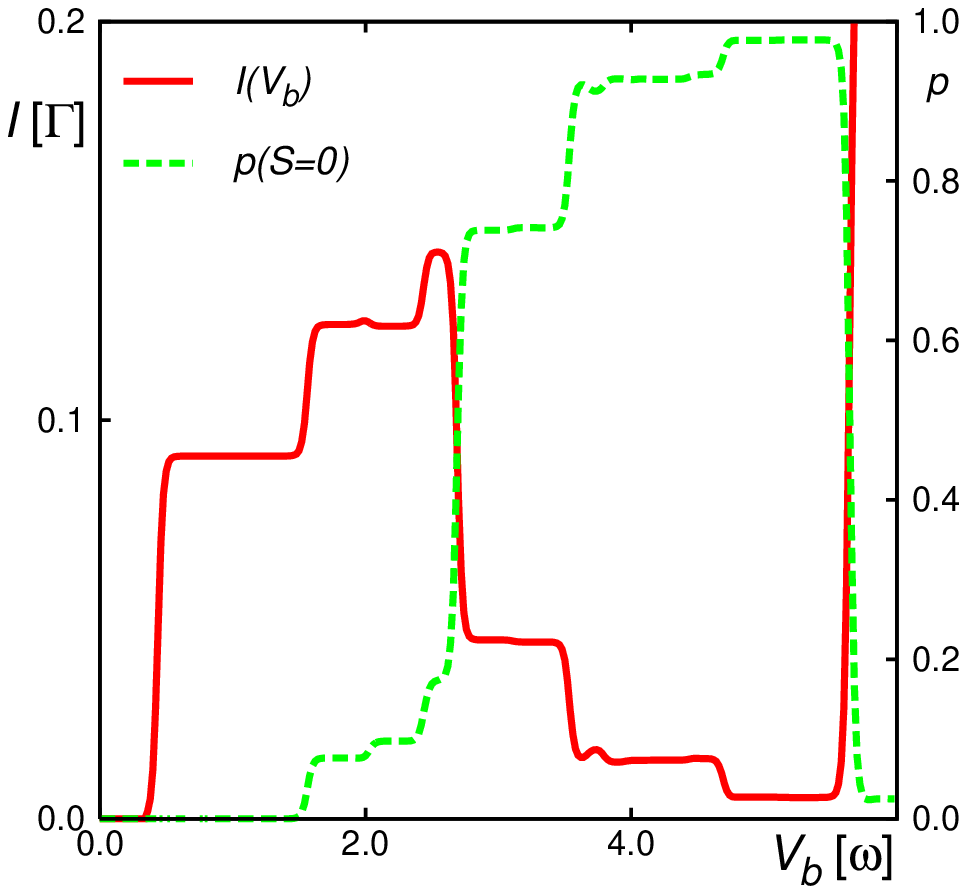}\\
      (b)\vspace{0.2cm}
    \end{tabular}\\
    \begin{tabular}{c}
      \includegraphics[width=5.5cm]{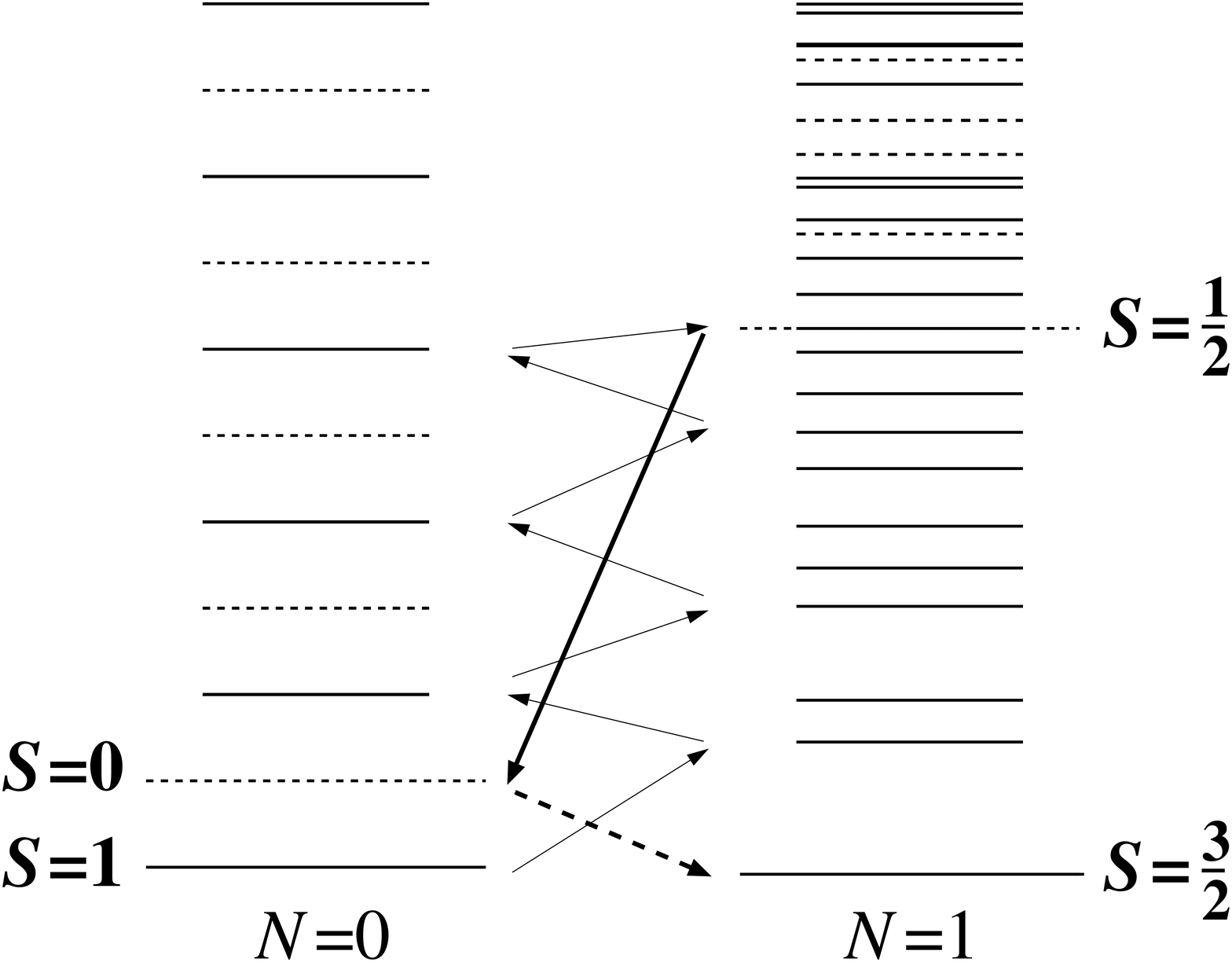}\\
      (c)
    \end{tabular}
  \end{tabular}
  \caption{
    (color online)
    (a) Differential conductance ($dI/dV_b$) for $V_g$ vs. $V_b$ ($J=0.5\omega$, $t=5.0\omega$, $\lambda=1.5$, $\Gamma=9\cdot10^{-4}\omega$, $\gamma^{\text{v}}_0=1.8\cdot10^{-2}\Gamma$, $\gamma^{\text{s}}_0=10^{-4}\Gamma$, $T=10^{-2}\omega$).
    (b) Current vs. $V_b$ (red solid) at $V_g=-0.2\omega$
    and probability
    of $S=0$ multiplet with no vibrational/vibronic quanta excited (green dashed).
    (c) The energy spectrum of the $N=0,1$ charge states and the spin-blockade mechanism.
    The line-style distinguishes the total spin values $S$ of the states.
    The longer lines denote a vibrational/vibronic groundstate
    whereas the shorter ones are excited by at least one such quanta.
    The molecule is ``pumped'' by a sequence of tunneling events,
    each time changing the charge and exciting vibrational and / or vibronic quanta
    until the $S=1/2$ spin state is reached.
    From there, the molecule falls into the blocking state ($S=0$
    vibrational groundstate) via another tunnel process.
    Since the transition to $S=3/2$ is forbidden by spin-selection rules,
    $S=0$ cannot relax and the current strongly suppressed.
  }
  \label{fig:result_blockade}
\end{figure}
For a wide range of parameters, the model exhibits a second, even more
striking effect, exemplified for $J=0.5\omega$, $t=5.0\omega$,
$\lambda=1.5$ in Fig.~\ref{fig:result_blockade}(a).
At low energies, the current is stepwise \textit{reduced}
when going deeper into the sequential tunneling region
leading to negative differential conductance (blue lines in Fig.~\ref{fig:result_blockade}(a)).
Simultaneously, the occupation of the molecular state with zero-spin and no vibrational excitations grows,
reaching over 90\% (see Fig.~\ref{fig:result_blockade}(b)).
Within this region the current is strongly suppressed
due to the pronounced population inversion that stabilizes the charge to $N=0$ and the spin to $S=0$.
This \emph{vibration-induced spin-blockade} provides another indication for the spin properties of the mixed-valence molecule
and additionally allows the spin to be controlled electrically.
The effect is readily understood by considering the non-equilibrium vibrations induced by the electric current.
First we note that in the low-bias region where the spin-blockade occurs,
the direct transition by electron
tunneling from $S=1 \rightarrow S=1/2$
is energetically not yet possible
and the transition $S=0 \leftrightarrow S=3/2$ is generally forbidden by spin selection rule ($\Delta S=\pm1/2$).
Now consider an electron which has just enough energy to excite
a vibrational, $Q_+$, (or vibronic, $Q_-$) quantum,
when entering / leaving the molecule ($N=0\leftrightarrow1$).
If the molecule does not immediately relax
it can accumulate more quanta in subsequent tunneling processes
as sketched in Fig.~\ref{fig:result_blockade}(c).
Eventually, when a sufficient amount of vibrational energy has been
accumulated a low energy electron can be assisted to excite the molecular spin-system.
This tunneling process brings the molecule to a lower spin state with $S=1/2$.
Finally, the molecule can relax to the $S=0$ state by a single tunneling process in which the excess energy is dissipated into the electrodes.
Now the molecule is trapped in a state with fixed charge $N=0$ and spin $S=0$
and the current is suppressed:
neither the $S=3/2$ states (due to the spin selection rule)
nor the $S=1/2$ state (due to the low bias voltage $V_b\lesssim (J + t)-2V_g$
and a strong Coulomb interaction on the molecule) are accessible.
Fig.~\ref{fig:result_blockade}(b) shows that the spin-blockade is lifted at higher bias $V_b\approx (J + t)-2V_g$
where the direct process back to $S=1/2$ becomes possible,
thereby confirming the above mechanism.
\par
Clearly, the vibration induced spin-blockade is expected to break down when
the excited spin state is too high in energy to be reached or when
escape processes from $S=0$ become dominant already at low bias voltages.
The escape processes, however, have to change the spin by \emph{at
least} 1 quantum and are therefore parametrically weak, since they
relate to spin-orbit coupling ($\gamma^{\text{s}}_0 \ll
\gamma^{\text{v}}_0$) or higher order tunneling processes.
For instance, the phenomenological spin-flip relaxation which we included
is responsible for the small remnant current in the spin-blockade region.
Secondly, the blocking state has to be reached: the single-electron transport current ``pumps''  the vibrational system (rate $\propto \Gamma$)
when the temperature is lowered below  the vibrational frequency $\omega$ (preventing thermal relaxation)
and when the tunnel coupling is sufficiently weak.
It is thus crucial that the intra-molecular vibrations are not strongly damped.
However,  relaxation rates  can compete with the transport rates
without destroying the vibration-induced spin-blockade
as long as the $S=1/2$ state can still be efficiently reached using the vibrations.
For the results we present here this is indeed the case.
Finally, cotunneling processes are expected to affect both the access to and
the escape from the blocking state.
A full calculation of this effect is possible in
principle~\cite{Leijnse08a}, but is prohibited by the large number of states
required here to describe the vibration-induced spin-blockade.
Inspection of the numerically calculated rates to second order in
$\Gamma$ for a truncated spectrum indicates that the presented results
are robust against perturbations due to higher-order tunneling.
Importantly, these processes can be suppressed by reducing the
tunneling coupling by appropriate choice of connecting ligand groups.
\par
We note that in this paper we have discussed the case of
ferromagnetic intra-molecular coupling $J$.
However, the vibration-induced spin-blockade is generic and also occurs for anti-ferromagnetic coupling $J<0$ provided that $t>|J|$
and in this case leads to a stabilized \emph{excess charge} $N=1$ and \emph{high spin} $S=3/2$.
\par
\section{Conclusions.}
Using a representative model
we have demonstrated that transport through a mixed-valence molecular transistor entails
an interplay of delocalized excess electrons, localized ionic spins,
ligand-shell vibrations and Coulomb blockade.
For this class of molecules transport-induced intra-molecular vibrations depend on the spin
and their energy can be transferred to the spin-system at specific voltages,
subsequently ``locking'' the spin, vibration and charge in a well-defined state.
The generic model, analyzed here in a non-equilibrium situation,
relates naturally to mixed-valence molecules~\cite{Bersuker92} such as Ru$^{2+,3+}$ complexes
with pyridine organic ligands of the 
Robin-Day class II or III.
The effects predicted here provide several bridges between nano-electromechanical systems (NEMS) and spintronic
devices and applications in this direction can be envisaged.
Clearly, the predicted spin-blockade effect will be sensitive to
local magnetic fields, mechanical energy dissipation and spin-orbit
effects and sensing applications involving these can be considered.
Also, the blockade effect indicates slow transport dynamics: this  may be
used in the context of switching and bistable operation of molecular
transistors by sweeping voltages non-adiabatically,
with the new possibility of magnetic field control due to the involvement of spin.
Finally, from a chemistry perspective, transport measurements provide unique insight in
the degree of electron delocalization determining the key properties of
mixed-valence molecules \emph{embedded} in an electric circuit.
Thus the investigation of complex mixed-valence systems as devices~\cite{Bominaar97}
proves to be an interesting avenue in single-molecule electronics.
\par
We acknowledge H. Schoeller, P. K{\"o}gerler, and H. Luecken for stimulating discussions
and the financial support from
 DFG SPP-1243,
 the NanoSci-ERA,
 the Helmholtz Foundation, and
 the FZ-J\"ulich (IFMIT).
\bibliographystyle{apsrev}

\end{document}